\documentclass[aps,pra,reprint,groupedaddress,nofootinbib,showkeys]{revtex4-2}

\usepackage{graphicx}
\usepackage[colorlinks=true,citecolor=blue,urlcolor=blue]{hyperref}

\begin{document}

\title{Properties of [111]-oriented ferroelectric superlattices \\
based on oxides with a perovskite structure}

\author{A. I. Lebedev}
\email[]{swan@scon155.phys.msu.ru}
\affiliation{Physics Department, Lomonosov Moscow State University, 119991 Moscow, Russia}


\begin{abstract}
The ground-state structures of fifteen free-standing [111]-oriented ferroelectric superlattices
(SLs) based on perovskite oxides and involving a single-atom substitution---BaTiO$_3$/SrTiO$_3$,
PbTiO$_3$/SrTiO$_3$, PbTiO$_3$/PbZrO$_3$, SrZrO$_3$/SrTiO$_3$, PbZrO$_3$/BaZrO$_3$, BaTiO$_3$/BaZrO$_3$,
PbTiO$_3$/BaTiO$_3$, BaTiO$_3$/CaTiO$_3$, KNbO$_3$/KTaO$_3$, KNbO$_3$/NaNbO$_3$, SrTiO$_3$/SrHfO$_3$,
BaTiO$_3$/BaSnO$_3$, PbTiO$_3$/CaTiO$_3$, SrTiO$_3$/CaTiO$_3$, and PbZrO$_3$/SrZrO$_3$---are calculated
from first principles. It is shown that in eleven considered SLs the ground state is ferroelectric.
A distinctive feature of the structure of the superlattices is that both the ferroelectric
distortions and the distortions associated with octahedral rotations are described by phonons
at the $\Gamma$~point of the Brillouin zone. The importance of taking into account structural
distortions associated with octahedral rotations, which in some cases lead to significant changes
in the ferroelectric properties of SLs, is demonstrated. The thermodynamic stability of superlattices
grown in the [111] and [100] directions is compared. An analysis of the force constants of the
superlattices helps to explain the peculiarities of the ferroelectric properties of the superlattices
and the corresponding solid solutions.
\end{abstract}


\widetext{\texttt{Author's translation of the paper published in Fizika Tverdogo Tela {\bf 68},
1242 (August 2026).}

\maketitle

\section{Introduction}

This work can be regarded as a continuation of paper~\cite{PhysSolidState.52.1448}, in which the
properties of ten [100]-oriented ferroelectric superlattices (SLs) based on oxides with a
perovskite structure were studied. In that paper, the ground-state structure of the SLs was
calculated from first principles within the density functional theory, and it was established that
the ground state of all studied superlattices is ferroelectric. The importance of taking into
account structural distortions caused by unstable phonons at the boundary of the Brillouin zone,
which in some cases led to significant changes in the ferroelectric and dielectric properties
of the superlattices, was demonstrated. In a subsequent paper~\cite{ComputMaterSci.188.110113},
the properties of [100]-oriented ferroelectric SLs with a polar discontinuity based on oxides
with a perovskite structure were studied.

In contrast to the relatively well-studied [100]-oriented superlattices, [111]-oriented
superlattices have been investigated to a much lesser extent. The most thoroughly studied
[111]-oriented SL is BaTiO$_3$/SrTiO$_3$~\cite{ApplPhysLett.77.3257,Vacuum.66.397,ApplPhysLett.103.112902}.
Similarly to [100]-oriented superlattices, short-period [111]-oriented SLs exhibit a higher
dielectric constant than films of the Ba$_{0.5}$Sr$_{0.5}$TiO$_3$ solid solution of the same
thickness and orientation. However, the remanent polarization ($2P_r = 0.027$~C/m$^2$) in these
SLs is noticeably lower. This latter property can be advantageous for microwave applications, as
lower polarization implies lower dielectric losses.

Among theoretical studies of [111]-oriented superlattices (SLs), the calculations presented
in Refs.~\cite{PhysRevB.59.12771,PhysRevB.63.092101,PhysRevB.65.224106,PhysRevB.66.214114,
PhysChemChemPhys.27.15920} are worth noting. These papers report results for the
PbTiO$_3$/PbZrO$_3$ superlattice. Information on the enthalpy of mixing in this superlattice
can also be found in Ref.~\cite{PhysSolidState.52.1448}. It should be noted that some authors
treat [111]-oriented short-period superlattices as a model for describing the properties of
disordered solid solutions. However, from the standpoint of the methodology used to construct
special quasi-random structures~\cite{PhysRevB.42.9622}, this choice is not optimal.

In Ref.~\cite{ChemMater.25.4545}, a series of [111]-oriented short-period superlattices
with substitution at the $A$~site was investigated theoretically. The main focus was on the
possible emergence of hybrid improper ferroelectricity driven by octahedral rotations in
structures with an ordered arrangement of atoms at the $A$~sites. It was found that
ferroelectricity can arise in such structures even in the absence of ions exhibiting a
pseudo-Jahn--Teller effect. However, the resulting structures were characterized by a very
high polarization-switching barrier, of the order of 1~eV, while the polarization values
were relatively small.

The aim of this work is to investigate the properties of free-standing, short-period
[111]-oriented ferroelectric superlattices based on perovskite oxides, with substitution
of only one atomic species on either the $A$ or $B$ sublattice. In
addition to the ten systems studied in Ref.~\cite{PhysSolidState.52.1448}, we consider
five other superlattices: BaTiO$_3$/BaSnO$_3$, SrTiO$_3$/SrHfO$_3$, PbTiO$_3$/CaTiO$_3$,
SrTiO$_3$/CaTiO$_3$, and PbZrO$_3$/SrZrO$_3$, some of which had previously been studied
only in the [100] orientation. The properties of these SLs are compared with those of
the same superlattices grown along the [100] direction. We do not consider superlattices
such as LaAlO$_3$/SrTiO$_3$, PbTiO$_3$/LaAlO$_3$, and BaTiO$_3$/BiFeO$_3$, which exhibit
a polar discontinuity, and we also do not investigate so-called double perovskites with
the $B$~site substitution, whose ordered (elpasolite) structure is analogous to that of
short-period superlattices with a [111] orientation. The obtained results may be used to
interpret data from future experiments on superlattices.

\section{Calculation technique}

In this work, we focus on the properties of short-period superlattices consisting of
alternating one-unit-cell-thick layers of two components that differ in the type of one
element. The crystal structure of the high-symmetry phase of such SLs is a
face-centered cubic (FCC) superstructure with NaCl-type ordering (space group
$Fm\bar{3}m$); when the substitution occurs at the $B$ sites, this corresponds to the
well-known elpasolite structure. In addition to SLs with the elpasolite structure,
we also examine SLs in which the substitution takes place at the $A$~sites of the
perovskite structure. For several systems (BaTiO$_3$/SrTiO$_3$, KNbO$_3$/KTaO$_3$,
BaTiO$_3$/BaZrO$_3$, BaTiO$_3$/BaSnO$_3$, and PbTiO$_3$/SrTiO$_3$), we also
investigate the properties of superlattices composed of a one-unit-cell-thick layer
of the first component and a layer of the second component that is two or three unit
cells thick. The symmetry of the high-symmetry phases in these SLs is described by
space groups $P\bar{3}m1$ and $R\bar{3}m$, respectively.

The calculations were performed from first principles within the density functional
theory, with pseudopotentials and wave functions expanded in a plane-wave basis, as
implemented in the \texttt{ABINIT} software package~\cite{abinit}. The
exchange-correlation interaction was described within the local density approximation
(LDA). Nonlocal optimized norm-conserving pseudopotentials in separable
form~\cite{PhysRevB.41.1227} were generated using the \texttt{OPIUM} software;
a local potential correction was also used to improve
transferability~\cite{PhysRevB.59.12471}. The plane-wave cutoff energy was 40~Ha
(1088~eV) for SLs containing Ta and Hf atoms, and 30~Ha (816~eV) for other SLs.%
    \footnote{In this work, several parameters used in the calculations are expressed
    in Hartree atomic units.}
Integration over the Brillouin zone was performed using a $4 \times 4 \times 4$ 
Monkhorst--Pack $k$-point mesh for 10-atom FCC unit cells and meshes with equivalent
$k$-point density for 20-atom cells. Atomic positions and lattice parameters were
relaxed until the Hellmann--Feynman forces fell below
$5 \times 10^{-6}$~Ha/Bohr (0.25~meV/{\AA}). Phonon spectra and elastic properties
were calculated using formulas derived from density-functional perturbation
theory (DFPT). The spontaneous polarization $P_s$ was computed using the Berry-phase
method.

\section{Results and their discussion}

\begin{figure*}
\centering
\begin{minipage}{.5\textwidth}
\includegraphics{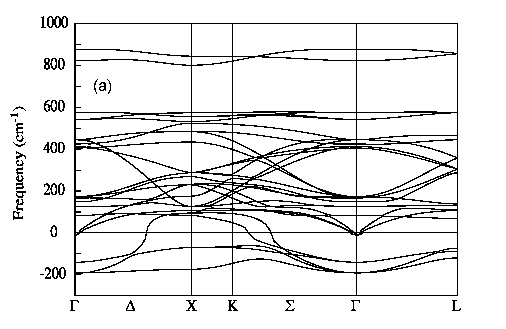}
\end{minipage}%
\begin{minipage}{.5\textwidth}
\includegraphics{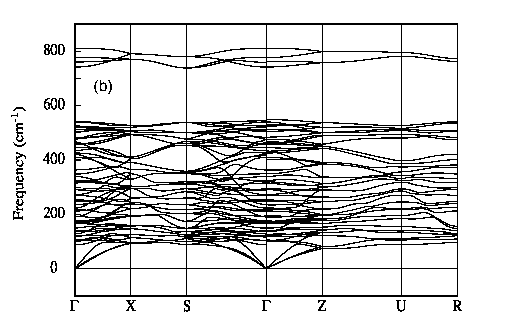}
\end{minipage}%
\caption{Phonon spectrum of a free-standing SrTiO$_3$/CaTiO$_3$ superlattice
($a$) in the paraelectric $Fm{\bar 3}m$ phase and ($b$) in the ground-state $Pmn2_1$
phase.}
\label{fig1}
\end{figure*}

The ground-state structure of a superlattice, characterized by the lowest total energy,
must exhibit positive frequencies for all optical phonons in the Brillouin zone and be
mechanically stable. Following the approach used in Ref.~\cite{PhysSolidState.52.1448},
the search for the ground state began by identifying a structure in which the frequencies
of all optical phonons at the center of the Brillouin zone were positive. To achieve this,
the equilibrium structure of the high-symmetry $Fm\bar{3}m$ phase was determined by
minimizing the Hellmann--Feynman forces, after which the phonon spectrum was calculated.
A typical phonon spectrum is shown in Fig.~\ref{fig1}(a). The frequencies of all unstable
phonons at the center of the Brillouin zone and at high-symmetry points on the zone boundary
for the $Fm\bar{3}m$ phase of all studied SLs are listed in Table~\ref{table1}. As follows
from the table, among the fifteen studied SLs, unstable phonons at the $\Gamma$~point,
indicated by imaginary mode frequencies, were observed in the $Fm\bar{3}m$ phase of all
SLs except BaTiO$_3$/BaZrO$_3$ and BaTiO$_3$/BaSnO$_3$. The symmetry of the unstable optical
phonons at the $\Gamma$~point depended on whether the $A$ or $B$ site of the $AB$O$_3$
perovskite structure was substituted during superlattice formation. For substitution at the
$A$~site, the unstable modes had $\Gamma_{15} (T_{1u})$ and $\Gamma_{25} (T_{2u})$ symmetry,
whereas for substitution at the $B$~site, the modes were $\Gamma_{15} (T_{1u})$ and
$\Gamma^{\prime}_{15} (T_{1g})$ (Table~\ref{table1}). The $\Gamma^{\prime}_{25} (T_{2g})$
mode appeared only in the PbTiO$_3$/PbZrO$_3$ SL and was so weak that it was not included
in our analysis. The $\Gamma_{15}$ mode is infrared active, the $\Gamma'_{25}$ mode is
Raman active, whereas the other two modes are silent.

It is well known that the $\Gamma_{15}$ mode describes ferroelectric structural distortions.
In [100]-oriented superlattices, the antiferrodistortive instability is described by phonons
at the $M$ and $R$~points of the Brillouin zone. In [111]-oriented superlattices, by contrast,
it is described by phonons at the $\Gamma$~point. These are the $\Gamma'_{15}$
mode, which corresponds to pure rotations, and the $\Gamma_{25}$ mode, which represents
rotations accompanied by strains. Therefore, to account for octahedral rotations in the
superlattices under study, there is no need to consider unit cells with doubled volume.

\begin{table*}
\caption{Frequencies of all unstable phonons or of the lowest-frequency stable phonon at
different points of the Brillouin zone for the paraelectric $Fm{\bar 3}m$ phase of studied
ferroelectric superlattices. The frequencies of the most unstable modes are shown in bold.}
\label{table1}
\begin{ruledtabular}
\begin{tabular}{cccccc}
Superlattice        & \multicolumn{5}{c}{Phonon frequencies, cm$^{-1}$} \\
\cline{2-6}
                    & $\Gamma$ & $L$ & $X$ & $W$ & $K$ \\
\hline
\multicolumn{6}{c}{Superlattices with substitution at the $A$~site} \\
BaTiO$_3$/SrTiO$_3$ & {\bf 101$i$} ($\Gamma_{15}$) & 88 & 76 & 91 & 83 \\
KNbO$_3$/NaNbO$_3$  & {\bf 191$i$} ($\Gamma_{15}$), 91$i$ ($\Gamma_{25}$)  & 46 & 127$i$, 78$i$, & 83$i$ & 79$i$, 10$i$ \\
                    &                                                      &                & 75$i$          &       & \\
PbTiO$_3$/SrTiO$_3$ & {\bf 121$i$} ($\Gamma_{15}$), 114$i$ ($\Gamma_{25}$) & 16             & 88$i$          & 31    & 15 \\
PbZrO$_3$/BaZrO$_3$ & {\bf 161$i$} ($\Gamma_{25}$), 112$i$ ($\Gamma_{15}$) & 118$i$, 89$i$, & 147$i$, 81$i$ & 117$i$, 90$i$, & 130$i$, 83$i$, \\
                    &                                                      & 60$i$          &               & 56$i$          & 72$i$, 27$i$ \\
PbTiO$_3$/BaTiO$_3$ & {\bf 140$i$} ($\Gamma_{15}$) & 26$i$ & 11 & 7 & 11$i$ \\
BaTiO$_3$/CaTiO$_3$ & {\bf 162$i$} ($\Gamma_{25}$), 150$i$ ($\Gamma_{15}$) & 116$i$, 99$i$, & 136$i$, 90$i$ & 106$i$, 95$i$, & 120$i$, 89$i$, \\
                    &                                                      & 54$i$          &               & 64$i$          & 79$i$, 45$i$ \\
PbTiO$_3$/CaTiO$_3$ & {\bf 187$i$} ($\Gamma_{25}$), 158$i$ ($\Gamma_{15}$) & 116$i$, 96$i$, & 173$i$, 80$i$ & 121$i$, 92$i$, & 145$i$, 80$i$, \\
                    &                                                      & 90$i$          &               & 51$i$          & 69$i$, 21$i$ \\
SrTiO$_3$/CaTiO$_3$ & {\bf 191$i$} ($\Gamma_{25}$), 140$i$ ($\Gamma_{15}$) & 120$i$, 89$i$, & 174$i$, 71$i$ & 118$i$, 85$i$, & 144$i$, 71$i$, \\
                    &                                                      & 73$i$          &               & 35$i$          & 59$i$ \\
PbZrO$_3$/SrZrO$_3$ & {\bf 195$i$} ($\Gamma_{25}$), 132$i$ ($\Gamma_{15}$) & 133$i$, 123$i$,  & 186$i$, 94$i$, & 138$i$, 131$i$, & 165$i$, 93$i$, \\
                    &                                                      & 120$i$, 102$i$,  & 39$i$, 32$i$   & 74$i$, 69$i$,   & 73$i$, 64$i$, \\
                    &                                                      & 30$i$            &                & 51$i$           & 52$i$, 42$i$, \\
                    &                                                      &                  &                &                 & 31$i$, 24$i$ \\
\multicolumn{6}{c}{Superlattices with substitution at the $B$~site} \\
KNbO$_3$/KTaO$_3$   & {\bf 104$i$} ($\Gamma_{15}$)                         & 133              & 30$i$          & 20              & 22 \\
PbTiO$_3$/PbZrO$_3$ & {\bf 161$i$} ($\Gamma^{\prime}_{15}$), 132$i$ ($\Gamma_{15}$), & 80$i$, 68$i$, & 153$i$, 64$i$, & 78$i$, 29$i$, & 124$i$, 59$i$, \\
                    & 5$i$ ($\Gamma^{\prime}_{25}$)                        &  55$i$           & 34$i$          & 18$i$, 10$i$, 3$i$ & 39$i$, 34$i$, 32$i$ \\
SrTiO$_3$/SrZrO$_3$ & {\bf 166$i$} ($\Gamma^{\prime}_{15}$), 55$i$ ($\Gamma_{15}$)  & 49$i$ & 148$i$ & 44$i$ & 114$i$ \\
BaTiO$_3$/BaZrO$_3$ &  97     &  92 & 94 & 95 & 93 \\
SrTiO$_3$/SrHfO$_3$ & {\bf 157$i$} ($\Gamma^{\prime}_{15}$) & 52 & 138$i$ & 49 & 96$i$ \\
BaTiO$_3$/BaSnO$_3$ & 77      &  93 &  93 &  93 & 92 \\
\end{tabular}
\end{ruledtabular}
\end{table*}

To search for the lowest-energy phase, small atomic displacements corresponding to the eigenvectors
of the identified unstable phonons were introduced into the $Fm\bar{3}m$ structure. Then, the
atomic positions and lattice parameters were relaxed to determine the equilibrium structures
of the lower-symmetry phases. Since the $\Gamma_{15}$, $\Gamma_{25}$, and $\Gamma'_{15}$ modes
are triply degenerate, it was necessary to consider all possible distortions permitted by a
three-component order parameter. In our analysis, we restricted ourselves to structures described
by the $(\eta,0,0)$, $(\eta,\eta,0)$, and $(\eta,\eta,\eta)$ order parameters. For the
$\Gamma_{15}$ phonon, the possible distorted structures are phases with the $I4mm$, $Imm2$,
and $R3m$ space groups; for the $\Gamma_{25}$ phonon, they are phases with the $I\bar{4}2m$,
$Imm2$, and $R32$ space groups; and for the $\Gamma'_{15}$ phonon, they are phases with the
$I4/m$, $C2/m$, and $R\bar{3}$ space groups. As an example, Table~\ref{table2} compares the
energies of different phases arising from the condensation of unstable phonons at the
$\Gamma$~point for the SrTiO$_3$/PbTiO$_3$ superlattice. For the resulting structures,
we checked whether any unstable phonons remained at the $\Gamma$~point. In the example under
consideration, all six phases mentioned above were found to be unstable. To search for a
lower-energy phase, atomic displacements corresponding to remaining unstable modes were
introduced into the unit cells, and the equilibrium structures were then determined by
relaxing the atomic positions and unit-cell parameters. The analysis of the phonon spectrum
and the search for new structures continued until a structure was found, in which the
frequencies of all optical phonons at the center of the Brillouin zone were positive. In
the example under consideration (Table~\ref{table2}), this phase turned out to be the $R3$
phase, which arose from the $R32$ and $R3m$ phases via the condensation of the corresponding
unstable $A_2$ modes. In superlattices with substitution at the $A$~site, the $R3$ phase
almost always possessed the lowest energy, with the only exception being the SrTiO$_3$/CaTiO$_3$
case. For brevity, we use the term intermediate ground state (IGS) to refer to identified
phases with no instability at the $\Gamma$~point. The symmetry of the IGS phases, which arise
from the condensation of all unstable phonons at the $\Gamma$~point, is presented in
Table~\ref{table3} for all superlattices under study.

\begin{table}
\caption{Energies of different phases (per 10-atom unit cell) arising from the
condensation of unstable phonons at the $\Gamma$~point in the SrTiO$_3$/PbTiO$_3$
superlattice.}
\label{table2}
\begin{ruledtabular}
\begin{tabular}{ccc}
Phase         & Phonon        & Energy (meV) \\
              & symmetry      & \\
\hline
$Fm{\bar 3}m$ &               & 0 \\
$R3m$         & $\Gamma_{15}$ & $-$38.35 \\
$I4mm$        & $\Gamma_{15}$ & $-$43.77 \\
$R32$         & $\Gamma_{25}$ & $-$46.39 \\
$I{\bar 4}2m$ & $\Gamma_{25}$ & $-$47.20 \\
$Imm2$        & $\Gamma_{15}$+$\Gamma_{25}$ & $-$56.38 \\
$R3$          & $R32$ + $A_2$ & $-$69.24 \\
\end{tabular}
\end{ruledtabular}
\end{table}

Among the superlattices with substitution at the $A$~site, in which the $Fm\bar{3}m$ phase
exhibited only a single ferroelectric $\Gamma_{15}$ unstable mode, there were
two---BaTiO$_3$/SrTiO$_3$ and PbTiO$_3$/BaTiO$_3$---for which the IGS phases were $R3m$ and $I4mm$,
respectively. In the other superlattices, both types of instabilities ($\Gamma_{15}$ and
$\Gamma_{25}$) were observed, leading to the same $R3$ structure with the exception of the
SrTiO$_3$/CaTiO$_3$ SL, for which the $Cm$ phase had the lowest energy.

\begin{table*}
\centering
\caption{Space groups and magnitudes of spontaneous polarization $P_s$ (in C/m$^2$) in the
intermediate ground state (IGS) and in the ground state (GS) of the studied [111]-oriented
free-standing ferroelectric superlattices. For comparison, the space groups and values of $P_s$
for the ground states of [100]-oriented superlattices are also provided \cite{PhysSolidState.52.1448}.}
\label{table3}
\begin{ruledtabular}
\begin{tabular}{ccccccc}
Superlattice        & \multicolumn{2}{c}{IGS [111]} & \multicolumn{2}{c}{GS [111]} & \multicolumn{2}{c}{GS [100]} \\
\cline{2-3}
\cline{4-5}
\cline{6-7}
                    & Space group & $P_s$           & Space group & $P_s$          & Space group & $P_s$ \\
\hline
\multicolumn{7}{c}{Superlattices with substitution at the $A$~site} \\
BaTiO$_3$/SrTiO$_3$ & $R3m$  & 0.253                & $R3m$          & 0.253       & $Cm$     & 0.241 \\
KNbO$_3$/NaNbO$_3$  & $R3$   & 0.489                & $R3$           & 0.489       & $Pmc2_1$ & 0.502 \\
PbTiO$_3$/SrTiO$_3$ & $R3$   & 0.436                & $R3$           & 0.436       & $Pmc2_1$ & 0.483 \\
PbZrO$_3$/BaZrO$_3$ & $R3$   & 0.378                & $R3$           & 0.378       & $Pmc2_1$ & 0.394 \\
PbTiO$_3$/BaTiO$_3$ & $I4mm$ & 0.581                & $I4mm$         & 0.581       & $Pmm2$   & 0.563 \\
BaTiO$_3$/CaTiO$_3$ & $R3$   & 0.473                & $R3$           & 0.473       & $Pc$     & 0.419 \\
PbTiO$_3$/CaTiO$_3$ & $R3$   & 0.582                & $Pc$           & 0.187       & $Pc$     & 0.560 \\
SrTiO$_3$/CaTiO$_3$ & $Cm$   & 0.256                & $Pmn2_1$       & 0.114       & $P1$     & 0.201 \\
PbZrO$_3$/SrZrO$_3$ & $R3$   & 0.407                & $Pmn2_1$       & 0.020       & $Pmc2_1$ & 0.416 \\
\multicolumn{7}{c}{Superlattices with substitution at the $B$~site} \\
KNbO$_3$/KTaO$_3$   & $R3m$  & 0.141                & $R3m$          & 0.141       & $Cm$     & 0.253 \\
PbTiO$_3$/PbZrO$_3$ & $R3$   & 0.620                & $R3$           & 0.620       & $Pc$     & 0.665 \\
SrTiO$_3$/SrZrO$_3$ & $C2/m$ & 0                    & $P2_1/c$       & 0           & $Pc$     & 0.218 \\
BaTiO$_3$/BaZrO$_3$ & $Fm{\bar 3}m$ & 0             & $Fm{\bar 3}m$  & 0           & $Pmc2_1$ & 0.259 \\
SrTiO$_3$/SrHfO$_3$ & $C2/m$ & 0                    & $P2_1/c$       & 0           & $P1$     & 0.117 \\
BaTiO$_3$/BaSnO$_3$ & $Fm{\bar 3}m$ & 0             & $Fm{\bar 3}m$  & 0           & $Pmc2_1$ & 0.214 \\
\end{tabular}
\end{ruledtabular}
\end{table*}

In superlattices with substitution at the $B$~site, the BaTiO$_3$/BaZrO$_3$ and
BaTiO$_3$/BaSnO$_3$ systems showed no instability at the $\Gamma$~point. In the
KTaO$_3$/KNbO$_3$ SL, only the $\Gamma_{15}$ ferroelectric mode was unstable, leading
to the lowest-energy $R3m$ phase. In the SrZrO$_3$/SrTiO$_3$ and SrTiO$_3$/SrHfO$_3$
SLs, the instability was determined solely by the $\Gamma^{\prime}_{15}$ mode and
resulted in the lowest-energy $C2/m$ phase. The most complex case was the
PbTiO$_3$/PbZrO$_3$ SL, in which distortions of the $\Gamma'_{15}$ symmetry produced
only a very small gain in energy, while the most influential unstable mode was
$\Gamma_{15}$. However, when the simultaneous instabilities of the $\Gamma_{15}$ and
$\Gamma'_{15}$ modes were taking into account, the $R3$ phase turned out to be the
lowest-energy phase.

For the identified IGS phases, the presence of other unstable modes was examined by
calculating phonon frequencies at high-symmetry points of the Brillouin zone. Such
unstable modes were observed quite frequently in the superlattices under consideration;
however, in only five of them did the resulting low-symmetry phases exhibit lower
energies than the corresponding IGS phases. For instance, in the SrTiO$_3$/SrZrO$_3$
and SrTiO$_3$/SrHfO$_3$ superlattices, the phonon spectra of the IGS phases ($C2/m$)
contained an unstable mode at the $M$~point, which led to a $P2_1/c$ phase.
Alternatively, the same result could be obtained by following the reverse sequence:
first, by accounting for the instability of the high-symmetry $Fm\bar{3}m$ phase
with respect to a phonon at the $X$~point, which leads to the $P4/mnc$ phase, and
subsequently by identifying the lowest-energy phase generated by an unstable $E_g$
phonon at the $\Gamma$~point of that phase. The resulting $P2_1/c$ phase satisfies
all stability criteria and is the ground state for the two superlattices in question.
In the $Fm\bar{3}m$ phase of PbTiO$_3$/CaTiO$_3$, SrTiO$_3$/CaTiO$_3$, and
PbZrO$_3$/SrZrO$_3$ superlattices, the initial instability of a phonon at the $X$~point
of the Brillouin zone ultimately results in the $Pmn2_1$ or $Pc$ phases, which are
the ground states of these superlattices (Table~\ref{table3}).

For all obtained IGS and GS structures, the elastic modulus tensors $C_{\mu\nu}$
were calculated, and the mechanical stability of the structures was verified
(this requires the determinant of the $6 \times 6$ elastic modulus matrix in Voigt
notation, as well as all its principal minors, to be positive). All the structures
were found to be mechanically stable. Thus, all identified GS structures with
the lowest energy satisfy all stability criteria and are the ground-state structures
(Table~\ref{table3}).

\begin{table*}
\caption{The relative difference in the volume of the unit cells of constituent
components in their ground states, $\Delta V/\bar{V}$, the space group, and the
values of the enthalpy of mixing, $\Delta H$, per formula unit for the ground
states of studied superlattices with [100] and [111] orientations.}
\label{table4}
\begin{ruledtabular}
\begin{tabular}{cccccc}
Superlattice       & $\Delta V/\bar{V}$, & \multicolumn{2}{c}{[100] SL}   & \multicolumn{2}{c}{[111] SL} \\
\cline{3-4}
\cline{5-6}
                   &      \%             & Space group & $\Delta H$ (meV) & Space group & $\Delta H$ (meV) \\
\hline
\multicolumn{6}{c}{Systems with substitution at the $A$~site} \\
BaTiO$_3$/SrTiO$_3$ &  7.51              & $Cm$     &  +20.0              & $R3m$    & +159.5 \\
KNbO$_3$/NaNbO$_3$  &  7.88              & $Pmc2_1$ &  +67.9              & $R3$     &  +71.4 \\
PbTiO$_3$/SrTiO$_3$ &  5.93              & $Pmc2_1$ &  +13.9              & $R3$     &  +20.1 \\
PbZrO$_3$/BaZrO$_3$ &  3.43              & $Pmc2_1$ &  +30.4              & $R3$     &  +27.4 \\
PbTiO$_3$/BaTiO$_3$ &  1.58              & $Pmm2$   &  +13.6              & $I4mm$   & +10.0 \\
BaTiO$_3$/CaTiO$_3$ & 14.21              & $Pc$     & +190.3              & $R3$     & +221.8 \\
PbTiO$_3$/CaTiO$_3$ & 12.64              & $Pc$     & +100.8              & $Pc$     & +114.8 \\
SrTiO$_3$/CaTiO$_3$ &  6.72              & $P1$     &  +62.9              & $Pmn2_1$ & +59.1 \\
PbZrO$_3$/SrZrO$_3$ &  3.54              & $Pmc2_1$ &  +15.4              & $Pmn2_1$ & +30.9 \\
\multicolumn{6}{c}{Systems with substitution at the $B$~site} \\
KNbO$_3$/KTaO$_3$   &  4.36              & $Cm$     &  +11.2              & $R3m$    &  +6.7 \\
PbTiO$_3$/PbZrO$_3$ & 11.36              & $Pc$     &  +43.1              & $R3$     & +42.1 \\
SrTiO$_3$/SrZrO$_3$ & 13.74              & $Pc$     &  +63.2              & $P2_1/c$ & +60.0 \\
BaTiO$_3$/BaZrO$_3$ & 13.20              & $Pmc2_1$ &  +41.5              & $Fm{\bar 3}m$ & $-$2.8 \\
SrTiO$_3$/SrHfO$_3$ & 11.03              & $P1$     &  +38.4              & $P2_1/c$ & +50.8 \\
BaTiO$_3$/BaSnO$_3$ & 14.03              & $Pmc2_1$ &  +31.0              & $Fm{\bar 3}m$ & $-$34.9 \\
\end{tabular}
\end{ruledtabular}
\end{table*}

The spontaneous polarization in the ground-state structures of all studied superlattices grown
along the [100] and [111] directions is compared in Table~\ref{table3}. Interestingly, the
ground states of all [100]-oriented SLs are polar, regardless of whether the constituent
materials are paraelectric, ferroelectric, or antiferroelectric. For [111]-oriented SLs, those
with substitution at the $A$~site are all polar, but only a few of those with substitution
at the $B$~site are polar. The polarization values for [100]-oriented SLs and for the IGS
phases of [111]-oriented SLs are comparable, although structural distortions in the
[100]-oriented SLs result in a noticeable reduction in spontaneous polarization. The fact
that the remanent polarization observed experimentally in the BaTiO$_3$/SrTiO$_3$
superlattice~\cite{ApplPhysLett.77.3257} is approximately 20~times lower than our calculated
value may be attributed to imperfections at the heterointerfaces or to the formation of
domain structures.

To evaluate the thermodynamic stability of the studied superlattices, we calculated
their enthalpies of mixing (per molecule) and compared the properties of [111]-oriented
SLs with those of [100]-oriented SLs. The enthalpy of mixing was determined as the
difference between the total ground-state energy of the SL under study and the arithmetic
mean of the total ground-state energies of its constituent components (note that in
previous similar calculations~\cite{PhysSolidState.52.1448}, the energies of the paraelectric
phases of the constituent components were used). The results of these calculations are
presented in Table~\ref{table4}. Since the studied SLs involve the substitution of only
one atomic species, the observed effects are ultimately connected with differences in the
size and electronic configuration of the substituted atoms. To estimate the size effect,
the table includes a column showing the relative difference in the unit-cell volumes (per
molecule) of the parent compounds in their ground states. The data reveal a clear correlation
between the enthalpy and the unit-cell volume difference. It is interesting that for SLs
involving substitution at the $B$~site, the enthalpies of mixing are notably lower thus
indicating that these structures accommodate the differences in atomic size more readily.

In addition to [111]-oriented superlattices with layer thicknesses of one monolayer,
the properties of superlattices with the second component having a thickness of two or
three monolayers were also investigated for several systems.  These superlattices showed
different behavior. In the (BaTiO$_3$)$_1$/(SrTiO$_3$)$_3$, (PbTiO$_3$)$_1$/(SrTiO$_3$)$_3$,
and (KNbO$_3$)$_1$/(KTaO$_3$)$_3$ superlattices, the ferroelectric instability was
preserved, consistent with the properties of the corresponding solid solutions. The
stability of the ferroelectric phase in the first two of these SLs can be understood
by noting that the --$B$--O--$B$--O-- chains, which are responsible for the emergence
of chain-like structural instability in niobates and
titanates~\cite{PhysRevLett.74.4067,Ferroelectrics.206.205}, remain intact. In contrast,
for the BaTiO$_3$/BaZrO$_3$ system, the ferroelectric instability was absent in the
(BaTiO$_3$)$_1$/(BaZrO$_3$)$_1$ superlattice and was observed only in the
(BaTiO$_3$)$_3$/(BaZrO$_3$)$_1$ SL. In the BaTiO$_3$/BaSnO$_3$ superlattice, the
ferroelectricity was absent even at a layer-thickness ratio of 3:1.

These results can be understood through an analysis of the interatomic force constants
matrices for the studied superlattices. Calculations of these matrices, the results of
which will be discussed in detail in a separate publication, show that in the
KNbO$_3$/KTaO$_3$ superlattice, the difference between Nb and Ta atoms is small;
consequently, the ferroelectric instability of the $B$O$_6$ cluster chains is preserved.
An analysis of the BaTiO$_3$/BaZrO$_3$ and BaTiO$_3$/BaSnO$_3$ superlattices reveals that
the larger Zr and Sn atoms tend to remain fixed at their lattice sites, and their
presence in the --$B$--O--$B$--O-- chains strongly suppresses the ferroelectric ordering.
Furthermore, it should be noted that the effective Born charge of the Sn atom
($Z^* = 4.601$) is only slightly higher than its nominal ionic charge, unlike the case
for Ti, Zr, Nb, and Ta atoms. This indicates a significant reduction in the covalent
character of the Sn--O chemical bond, which is directly linked to the electronic
contribution to the ferroelectric instability.

In the PbTiO$_3$/PbZrO$_3$ superlattice, the perturbation of the --$B$--O--$B$--O--
chain, as in its barium counterpart, is significant and effectively suppresses
the ferroelectric instability of the chain. However, in this superlattice, as in pure
PbTiO$_3$, there is a strong interaction within the chains of PbO$_{12}$ clusters. This
interaction preserves the ferroelectric instability of the structure, both in the
superlattice and across a wide range of the corresponding solid solutions. Distortions
in the Pb chains produced by the partial substitution of Pb with Sr or Ba sharply weaken
interatomic interactions, thus explaining the rapid decrease in the phase-transition
temperature observed in solid solutions. Interestingly, the calculations revealed that
Pb atoms in PbTiO$_3$ and superlattices derived from it do not exhibit off-center
displacement, a feature often cited as the origin of the ferroelectric instability
in these materials, which is attributed to the stereochemical activity of the lone
$s^2$ electron pair~\cite{JChemSoc.1959.3815}.

\section{Conclusion}

In this work, the ground-state structures of fifteen [111]-oriented ferroelectric
superlattices (SLs) based on oxides with a perovskite structure (BaTiO$_3$/SrTiO$_3$,
PbTiO$_3$/SrTiO$_3$, PbTiO$_3$/PbZrO$_3$, SrZrO$_3$/SrTiO$_3$,
PbZrO$_3$/BaZrO$_3$, PbZrO$_3$/SrZrO$_3$, BaTiO$_3$/BaZrO$_3$,
PbTiO$_3$/BaTiO$_3$, BaTiO$_3$/CaTiO$_3$, KNbO$_3$/KTaO$_3$, KNbO$_3$/NaNbO$_3$,
SrTiO$_3$/SrHfO$_3$, BaTiO$_3$/BaSnO$_3$, PbTiO$_3$/CaTiO$_3$, and
SrTiO$_3$/CaTiO$_3$) were determined from first-principles calculations. It was shown
that all studied superlattices with the substitution at the
$A$~site exhibit a ferroelectric ground state, whereas this behavior is less common for
superlattices with substitution at the $B$~site. The polarization of the ground state
and the enthalpy of mixing were calculated for all SLs. The importance of accounting
for non-ferroelectric structural distortions, which in some cases lead to significant
changes in ferroelectric properties, is demonstrated. An analysis of the interatomic
force constants in the superlattices helps to explain the observed trends in the
ferroelectric properties of both the SLs and the corresponding solid solutions.


\providecommand{\BIBYu}{Yu}

\end{document}